\documentclass{article}
\title{\bf 
Thin Fisher Zeroes
}
\author{ {\it B. P. Dolan}\\
Department of Mathematical Physics,\\
National University of Ireland,\\
Maynooth, Ireland\\
and\\
School of Theoretical Physics\\
Dublin Institute for Advanced Studies\\
10, Burlington Rd. \\
Dublin, Ireland\\ 
\\
{\it W. Janke}\\
Institut f\"ur Theoretische Physik,\\
Universit\"at Leipzig,\\
Augustusplatz 10/11,\\
D-04109 Leipzig, Germany\\
\\
{\bf and}\\
\\
{\it D.A. Johnston} and {\it M. Stathakopoulos}\\
         Dept. of Mathematics\\
         Heriot-Watt University\\
         Riccarton\\
         Edinburgh, EH14 4AS, Scotland
         }
\begin{document}
  \maketitle
                      {\Large
                      \begin{abstract}
%
Biskup {\em et al.\/} [Phys. Rev. Lett. {\bf 84} (2000) 4794] have recently suggested that the loci of partition function zeroes 
can profitably be regarded as phase boundaries in the complex temperature or field
planes. We obtain the Fisher zeroes for Ising and Potts models on non-planar (``thin'')
regular random graphs using this approach, and note that the locus of Fisher zeroes
on a Bethe lattice is identical to the corresponding random graph. Since
the number of states $q$ appears as a parameter in the Potts solution
the limiting locus of chromatic zeroes is also accessible.
%
                        \end{abstract} }
%
  \thispagestyle{empty}
%
%
  \newpage
%
                  \pagenumbering{arabic}

\section{Introduction}

The idea that studying the zeroes of the partition function
of a statistical mechanical model when a physical parameter was
extended to complex values might provide valuable insights
on critical behaviour in lattice models was first promulgated by
Lee and Yang \cite{YL} for field-driven transitions and later by Fisher
and others for temperature-driven transitions \cite{Fish,IPZ}.
Although there has been a considerable body
of work studying the partition function zeroes of the Ising and Potts
model on various regular lattices, most notably
in recent years in  \cite{lottsashrock,creswick}, there has been
little investigation of the spin models on lattices exhibiting some
form of geometrical disorder, such as random graphs, both for the annealed 
and quenched case. Some cases that
have been investigated such as Penrose tilings have revealed intricate
structure for the Fisher (complex temperature)
zeroes away from the real axes, while still possessing Onsager
scaling behaviour at the physical critical point \cite{prz}. Others,
such as the Ising model on annealed ensembles of planar $\phi^4$ random graphs
and the dual quadrangulations possess Fisher zeroes which do lie
on curves in the complex temperature plane
\cite{jan}, but which are different from
the double-circle locus of the regular square lattice. This in itself
is interesting, since it shows that a distinct locus still exists when
a sum over a class of graphs is folded into the partition function along
with a sum over spin configurations. For both planar%
\footnote{We denote 
the planar graphs as ``fat'' graphs since they can be thought of as being
generated by the Feynman diagrams in a matrix theory which has fat, or
ribbon-like, propagators that carry one matrix index on each side.
The fatness of the propagators allows the identification of a surface
with a given graph and helps enforcing planarity.
The converse is that generic random graphs appear as the Feynman diagrams
in a theory with ``thin'' propagators that carry no edge labels and
which have no surface interpretation.}
and generic, 
non-planar random graphs perturbative calculations strongly suggest
that the Lee-Yang (complex field) zeroes still lie on the unit circle in
the complex activity plane \cite{jan,brazil}, 
even though the technical assumptions of the original Lee-Yang theorem 
no longer apply in this case.

The previous work on both fat and thin random graph models has relied 
entirely  on perturbative
expansions and Monte-Carlo simulations to investigate the various loci of 
zeroes. There is rather more information at hand in both these cases, however,
since exact evaluations of the partition function are possible 
in the thermodynamic limit using
matrix integral/orthogonal polynomial methods in the planar case and
a straightforward  saddle-point calculation in the non-planar case.
In the case of the Ising model on regular two-dimensional lattices
the particular form of the solution
for the free energy, as a double trigonometrical integral over
a logarithm, allowed a direct identification of the loci of zeroes
as the curves or regions where the argument of the logarithm was zero.
The fat and thin random graph solutions no longer have this form --
they are even simpler, being directly the logarithms of algebraic expressions,
which differ in the high and low temperature phases but match
at the transition point, as they should.

The general question of how to extract loci of zeroes 
for models with first order transitions when
one has some analytical expression or approximation
for the partition function
or free energy available 
has recently been addressed by 
Biskup {\em et al.\/} \cite{BK}. They showed that under suitable technical conditions
the partition function of a $d$-dimensional 
statistical mechanical
model defined in a periodic volume $V = L^d$ which depends
on a complex parameter $z$ can be written in terms of
complex functions $f_l(z)$ describing $k$ different phases as
\begin{equation}
Z = \sum_l^k q_l e^{ - \beta f_l V} +  O ( e^{ - L/ L_0} e^{- \beta f V} )
\end{equation}
where $q_l$ is the degeneracy of phase $l$, $\beta$ is the inverse temperature
and $L_0$ is of the order of the correlation length.
The various $f_l$ are to be interpreted as the metastable 
free energies of the phases, with $\Re f_l = f$ characterising
the free energy when phase $l$ is stable.
The zeroes of the partition function are then determined 
to lie within $O ( e^{ - L/ L_0 } )$ of the solutions of the
remarkably simple equations
\begin{eqnarray}
\Re f_{l,L}^{eff} =  \Re f_{m,L}^{eff} < \Re f_{k,L}^{eff}, \; \; \forall k \ne l,m \nonumber \\
\beta V ( \Im  f_{l,L}  -  \Im   f_{m,L} ) = \pi \; mod \; 2 \pi
\label{master}
\end{eqnarray}
where $f_l^{eff} = f_l - ( \beta V )^{-1} \log q_l$.
Since we have written an expression for $Z$ that incorporates phase
co-existence we are implicitly looking at models with first
order transitions, the canonical example being the field-driven transition
for the Ising model.

The equations (\ref{master}) show that
zeroes of $Z$ asymptotically lie on the complex phase coexistence
curves $\Re f_{l,L} = \Re f_{m,L}$ in the complex $z$-plane.
Indeed, the idea that the loci of zeroes might be thought of as Stokes lines
which separate different asymptotic behaviours and that the real part of 
the free energy in the various phases should match along the lines is
already implicit in \cite{IPZ}.
 
It might appear erroneous to attempt
to apply these results, formulated for
first order transitions and based on ideas arising from phase
co-existence, to models where the physical 
temperature-driven transition is continuous.
However, when considered 
in the complex temperature plane 
such transitions will be continuous only at the 
physical point itself and possibly some other
finite set of points. This can be seen by looking at expressions for
the magnetization for the Ising model, for example, on the square lattice,
on fat $\phi^4$ graphs and on thin $\phi^3$ graphs: 
\begin{eqnarray}
M &=& { ( 1 + u)^{1/4} ( 1 - 6 u+ u^2 )^{1/8} \over (1 - u)^{1/2} } \nonumber \\
M &=& { 3 ( 1 - 16 c^2 )^{1/2} \over 3 - 8 c^2 } \\
M &=&   {( 1 - 3 c )^{1/2} \over (1 - 2 c  ) ( 1 + c )^{1/2}} \nonumber
\end{eqnarray}
where $u=c^2=\exp(-4\beta)$.
The expressions for $M$ apply through the complex extension of the low-temperature phase, and $M$ will be zero outside this region. We can thus see that although $M$
will vanish continuously at the physical critical points, $u=3 -2 \sqrt{2}, \;
c= 1/4, \; c = 1/3$ respectively, 
it will generically be non-zero at the phase boundary
approaching from within the low temperature region, whereas it will be zero
approaching from outside. 
There are further discrete points where the magnetization  
vanishes continuously: at the anti-ferromagnetic point $u=3 +2 \sqrt{2}$
and the unphysical point $u=-1$ on the square lattice; and the
unphysical point $c=-1/4$ on the planar $\phi^4$ graphs. It should also be remarked that
the apparent singularities in the
expressions for $M$ at $u=1$, $c = \sqrt{3/8}$ and $c=1/2, -1$ respectively
are spurious since they lie outside the domain of applicability of the expressions.
Since we only have a discrete set of points along
the phase boundary where $M$ vanishes continuously, a typical point 
will be first order in nature and
the arguments of \cite{BK} may safely
be deployed to determine the locus of zeroes.

The results
of \cite{BK}
give a very natural interpretation of the loci of partition function 
zeroes as phase coexistence curves for suitably complex extended phases,
both for Lee-Yang and Fisher zeroes.
Such a picture is also 
in accordance with the ``dynamical systems'' determination
of partition function zeroes \cite{various} for models which may be defined
in terms of some sort of recursive renormalization group map. In these cases
the loci of zeroes appear as the Julia set of the map -- in other words, the
boundaries of the basins of attraction for the map, or phase boundaries
of the complex extended phases. 

In this paper we find the locus of Fisher
zeroes for  the Ising model on 
3-regular and 4-regular non-planar 
random graphs \cite{Bac,dj}, 
where such an expression can be calculated by saddle point methods
and on the corresponding Bethe lattices where rather different recursive
methods arrive at an identical answer. 
As a warm up exercise we show that the methods employed in the study
of 3-regular and 4-regular graphs also reproduce the known results 
for zeroes that have been obtained by transfer
matrix methods \cite{Glu} in the one-dimensional 
\footnote{i.e. on a 2-regular graph.} Ising model.
The $q$-state Potts model on 3-regular and 4-regular random lattices
may be treated in a similar manner
to the Ising model and the locus of Fisher zeroes obtained for general
$q$. Since $q$ appears in the solution for the $q$ state Potts model 
as a parameter we are also at liberty to look at chromatic zeroes 
for our random graphs, which involves looking at the saddle point solutions
for the ground state of the anti-ferromagnetic Potts model in the complex
$q$-plane.

\section{Ising model in one dimension}

The partition function for the Ising model
with Hamiltonian
\begin{equation}
-\beta H = \beta \sum_{<ij>} \sigma_i \sigma_j,
\label{isingH}
\end{equation}             
where $\sigma_{i} = \pm 1$,
on 2-regular random graphs
with $n$ vertices (i.e. on a one-dimensional ring)
may be written as
\cite{Bac,dj}
\begin{equation}
Z_n(\beta) \times N_n = {1 \over 2 \pi i} \oint { d g \over
g^{n + 1}} \int { {d \phi_+ d \phi_- \over 2 \pi \sqrt{\det K}}
\exp (- S )},
\label{part2}
\end{equation}
where $N_n$ is the number of undecorated rings and 
$K$ is defined by
\begin{equation}
\begin{array}{cc} K_{ab}^{-1} = & \left(\begin{array}{cc}
1 & -c \\
-c & 1
\end{array} \right) \end{array}.
\end{equation}
The action itself is
a direct transcription of the
matrix model action \cite{boul} to scalar variables
\begin{equation}
S = {1 \over 2 } \sum_{a,b}  \phi_a  K^{-1}_{ab} \phi_b  -
{g \over 2} ({ z  } \phi_+^2 + { 1 \over  z }\phi_-^2).
\label{e2}
\end{equation} 
with $z = \exp ( h )$ and
$c = \exp ( - 2 \beta )$ .
In the above $\phi_+^2 $ can be thought of as representing ``up'' spin
vertices
and  $\phi_-^2 $ ``down'' spin vertices. Since the integral is gaussian
we are not obliged to pursue the saddle-point evaluation that is necessary
in the next section for 3-regular and 4-regular graphs -- we can simply carry out the 
integral directly. Scaling away the vertex counting factor of $g$
we find that the free energy per site is given by the logarithm
of the dominant eigenvalue of the quadratic form in the propagator,
\begin{eqnarray}
f &\sim& \log ( \lambda_\pm )  \nonumber \\
\lambda_\pm &=& { - ( 1 - z)^2 \pm \sqrt{ ( 1 - z^2)^2 + 4 c^2 z^2 } \over 2 z }.
\end{eqnarray} 
The partition function zeroes and hence free energy singularities
appear when the square root 
in the above expression is zero and both roots can contribute
\begin{equation}
( 1 - z^2)^2 + 4 c^2 z^2 = 0
\end{equation}
which recovers the equation for the Lee-Yang edge derived in \cite{Glu}.

\section{Ising model on thin random graphs}

The partition function for the Ising model
with the Hamiltonian of equ.~(\ref{isingH})
on 3-regular random graphs
with $2n$ vertices may be written as
\cite{Bac,dj}
\begin{equation}
Z_n(\beta) \times N_n = {1 \over 2 \pi i} \oint { d g \over
g^{2n + 1}} \int { {d \phi_+ d \phi_- \over 2 \pi \sqrt{\det K}}
\exp (- S )},
\label{part}
\end{equation}
where $N_n$ is now the number of undecorated (no spin) $\phi^3$ graphs.
$K$ is defined as above and 
the action now has the appropriate
potential for generating $\phi^3$ graphs
\begin{equation}
S = {1 \over 2 } \sum_{a,b}  \phi_a  K^{-1}_{ab} \phi_b  -
{g \over 3} (\phi_+^3 + \phi_-^3).
\label{e3}
\end{equation}
It is necessary to include the counting factor $N_n$ to disentangle
the factorial growth of the undecorated graphs from any
non-analyticity due to phase transitions in the decorating spins.
One is also obliged to pick out the $2n$-th order in the expansion
explicitly with the contour integral over $g$ as, unlike
the planar graphs of two-dimensional gravity,
$g$ cannot be tuned to a critical value to cause a divergence. 

The mean-field Ising transition manifests itself in this formalism
as an exchange of dominant saddle points. Solving the
saddle point equations $\partial S / \partial \phi_{\pm} = 0$
\begin{eqnarray}
\phi_+ &=& \phi_+^2 + c \phi_-, \nonumber \\
\phi_- &=& \phi_-^2 + c \phi_+,
\end{eqnarray}
which we have rescaled to remove $g$  and an irrelevant
overall factor, we find a symmetric high-temperature solution
\begin{equation}
\phi_+ = \phi_- = 1 - c,
\label{isingsol0}
\end{equation}
which bifurcates at $c=1/3$ to the low-temperature
solutions
\begin{eqnarray}
\phi_+ &=& { 1 + c + \sqrt{1- 2 c - 3 c^2} \over 2 }, \nonumber \\
\phi_- &=& { 1 + c - \sqrt{1- 2 c - 3 c^2 } \over 2}.
\label{isingsol}
\end{eqnarray}
The bifurcation point is determined by the value of $c$ at which
the high- and low-temperature solutions for $\phi$ are identical,
which appears at the zero of the Hessian $\det ( \partial^2 S / \partial \phi^2
)$.
The magnetisation order parameter for the Ising model
can also be transcribed directly from the matrix model \cite{boul}
\begin{equation}
M = { \phi_+^3 - \phi_-^3 \over \phi_+^3 + \phi_-^3}
\end{equation}
and shows a continuous transition with mean-field
critical exponent ($\beta=1/2$). The other critical 
exponents may also be calculated
and take on mean-field values.

If one carries out a saddle point evaluation of equ.~(\ref{part}), the 
leading term in the free
energy will be given by an expression of the form
$f \sim \log \tilde S$, where $\tilde S$ is the action (\ref{e3})
evaluated at either the low- or high-temperature saddle point solutions
for $\phi$ in eqs.~(\ref{isingsol0},\ref{isingsol}). Since we wish
to match $\Re f$ between phases in order to find the loci of the partition function zeroes we are demanding in this case
\begin{equation}
| \tilde S_L | = | \tilde S_H |,
\label{match}
\end{equation}
where $S_L$ is the low-temperature saddle-point solution
and $S_H$ is the high-temperature saddle-point solution,
both to be taken for complex $c$
\begin{eqnarray}
S_H &=& \frac{1}{3} (1-c)^3, \nonumber \\ 
S_L &=&  \frac{1}{6} (1 + c)^2 ( 1 - 2 c).
\end{eqnarray} 

The resulting locus of Fisher zeroes is shown in Fig.~1, and is slightly unusual
in that it is not a closed curve as is usually the case for Ising and Potts
models on regular lattices, but rather a cusp. It is not
unique in not being a closed curve, however, since the 
the locus
of the Fisher zeroes for the one-dimensional Ising model
is all of the imaginary axis \footnote{The one-dimensional
locus of Fisher zeroes can be found by applying the field/temperature
duality unique to the one-dimensional case to the Lee-Yang zeroes
on the unit circle.}. For one-dimensional Potts
models a similar result holds, with the zeroes lying on a vertical
line in the $\exp ( - 2 \beta )$ plane passing through the real axis
at $1 - q/2$.
The random graph Ising model has mean-field exponents because of the
locally tree-like structure, so the scaling relation relating the 
angle $\psi$ which the zeroes make as they approach the real axis,
the critical heat exponent $\alpha$ and the two 
critical amplitudes $A_+, A_-$,
\begin{equation}
\tan \left[ ( 2 - \alpha) \psi \right] = { \cos ( \pi \alpha ) - A_+ / A_- \over \sin ( \pi \alpha ) },
\label{slope}
\end{equation}
predicts%
\footnote{$\alpha=0, A_+/A_-=0$. Note that the original discussion was in terms
of $u=c^2$, but since we are dealing with conformal transformations any conclusions about
angles will still hold, as can easily be confirmed by plotting the locus directly
in terms of $u$.} 
$\psi = \pi / 4$.
This is indeed what is observed at the cusp of
the locus ($c = 1/3$) in Fig.1. 

It should be remarked that a little care is needed
in applying equ.~(\ref{match}) since  the presence of the moduli signs means
that one runs of  risk of picking up ``wrong sign'' solutions.
This is easily resolved when the complex phases are extensions of real phases
as in the present case, when the behaviour on the real axis can be examined 
to make sure that the correct sign has been chosen.

The determination of the Fisher zeroes locus on other random regular graphs
proceeds in the same manner as on the $\phi^3$ graphs. On
$\phi^4$ (four-regular) graphs, for instance the action becomes
\begin{equation}
S = {1 \over 2 } \sum_{a,b}  \phi_a  K^{-1}_{ab} \phi_b  -
{g \over 4} (\phi_+^4 + \phi_-^4),  
\end{equation}
and the saddle-point solutions are 
\begin{eqnarray}
\phi_+ = \phi_- = \sqrt{1-c}, 
\end{eqnarray}
and
\begin{eqnarray}
\phi_+ &=&  \frac{1}{4 c}(\sqrt{2 + 2 \sqrt{1 - 4 c^2}}) ( 1 - \sqrt{1 - 4 c^2}), \nonumber \\  
\phi_- &=& \frac{1}{2} ( \sqrt{2 + 2 \sqrt{1 - 4 c^2}} )
\end{eqnarray}
at high and low temperatures, respectively. These lead to the 
saddle point actions
\begin{eqnarray}
S_H &=& -c + \frac{1}{2} + \frac{1}{2} c^2, \nonumber \\
S_L &=& \frac{1}{4} - \frac{1}{2} c^2,
\end{eqnarray}
which are matched in the same manner as for the $\phi^3$
graphs and also lead to a cusp locus, with the cusp now at $c=1/2$.

Both the $\phi^3$ and $\phi^4$ curves appear rather like a Folium of Descartes
with the loop that would lie to the right of the cusp excised. The $\phi^3$
curve turns out to have a rather more complicated formula
\begin{equation}
\frac{27}{16} y^2 = { -(1 + \frac{9}{2}x + \frac{27}{16}x^2 + \frac{27}{4}x^3) 
+ (1 + 3 x)^{3/2} \over
 (1+4x) }
\end{equation}
which is
deduced from expressing $|S_H| = |S_L|$ in terms of the real, $x = \Re v$, and
imaginary, $y = \Im v$ parts of $v=c - 1/3$. 
However, the expression for the $\phi^4$ curve
\begin{equation}
y^2= {-x^2(x-\frac{1}{4}) \over (x+\frac{1}{4})}
\end{equation}
where $x = \Re v$, $y = \Im v$ and $v=c - 1/2$ is indeed a rescaling
of the standard Folium.

\section{Ising model on Bethe lattice}

We now turn to the Ising model on a Bethe lattice, with co-ordination number $m$.
The free energy per site is arrived at in an apparently very different manner
to the random graphs
by considering a recursion relation on the branches of the tree and 
is given by
\begin{equation}
 \beta f =- {1\over 2}m \beta - {1\over 2} m \ln (1 - c^2)
+ {1\over 2} \ln [c^2+1 - c(x+x^{-1})]
+{1\over 2}(m - 2) \ln (x + x^{-1} - 2c),
\label{treepart} 
\end{equation}
where $c$ is again equal to
$\exp(-2 \beta)$ and $x$, defined implicitly by              
\begin{equation}
x = {{e^{-2 \beta} + e^{-2h} x^{m-1}}\over{1 + e^{-2h-2 \beta} x^{m-1}}},
\label{curses}
\end{equation}
emerges as the fixed point of the recursion. 
This model has a phase transition at $\beta_c =
{1\over 2} \ln [m/ ( m-2 )]$ and
the magnetisation per site is
\begin{equation}
M = {{e^{2h} - x ^m}\over{{e^{2h} + x^m}}},
\end{equation}
which ranges from $-1$ for $x \rightarrow \infty$ to $+1$ for
$x = 0$. The derivation of these equations 
can be found in \cite{Bax}.

If we restrict ourselves to $h=0$ and $m=3$ in order 
to compare with the Fisher zeroes on $\phi^3$ graphs, the
solutions of equ.~(\ref{curses}) are given by
\begin{eqnarray}
x &=&  1, \nonumber \\
x &=&  { 1 - c \pm \sqrt{ 1 - 2 c - 3 c^2 } \over 2 c},
\end{eqnarray}
with $x=1$ being a high-temperature solution (since $M=0$ in this case)
and the others being low-temperature solutions when $T<T_c$.
If we substitute  these into equ.~(\ref{treepart}) we find
\begin{eqnarray}
\beta f_H &=&  - \frac{3}{2} \beta - \frac{3}{2} \log ( 1 - c^2)
+ \frac{1}{2} \log [ 2 ( 1 - c)^3 ],  \nonumber \\
\beta f_L &=&  - \frac{3}{2} \beta - \frac{3}{2} \log ( 1 - c^2)  
+ \frac{1}{2} \log [( 1 + c )^2 ( 1 - 2 c) ],
\label{betaf}
\end{eqnarray}
where we have deliberately kept the terms that do not depend on
$x$, and hence are the same in both the high- and low-temperature phases,
separate. It is clear from these expressions that demanding
$\Re f_L = \Re f_H$ gives us back exactly the same expression for the
Fisher locus as
in the random graph case, namely:
\begin{equation}
| 2 ( 1 - c)^3 | = | ( 1 + c )^2 ( 1 - 2 c) |.
\end{equation}
The first two terms in each of eqs.~(\ref{betaf}) have been spirited away
into the similarity sign 
in $f \sim \log \tilde S$ in the random graph expressions,
where they arise from the normalising $\det K$ in equ.~(\ref{part}).

That the Bethe lattice with three neighbours should have the same
Fisher zeroes as the $\phi^3$ random graph is not unexpected, since
the ratio of the saddle point solutions for the random graph model
gives precisely the recursion relation (\ref{curses})
and both exhibit mean-field critical behaviour. The
fixed point of the recursion variable $x$ in the Bethe lattice solution
is equal to the ratio of the $\phi_+$ and $\phi_-$ solutions in the random 
graph case. 
This equivalence between the random graph and Bethe lattice models 
is quite generic and also applies to the Potts models which
we now move on to discuss.

\section{Potts model on thin random graphs}

A similar expression to equ.~(\ref{part}) can be used
to define the partition function for a $q$-state Potts model
where the
Hamiltonian can be written
\begin{equation}
-\beta H =   \beta \sum_{<ij>} ( \delta_{\sigma_i, \sigma_j} -1).
\end{equation}
The spins $\sigma_i$ now take on $q$ values, so
one might expect by analogy with the 
Ising model ($q=2$) that $q$ ``fields'' $\phi$ might be needed.
This leads to a Potts action of the form \cite{potts}
\begin{equation}
S = { 1 \over 2 } \sum_{i=1}^{q} \phi_i^2 - c \sum_{i<j} \phi_i \phi_j -{1 \over
 3} \sum_{i=1}^q \phi_i^3,
\label{qstate}
\end{equation}   
with  $c = 1/ ( \exp( 2 \beta)  + q-2)$.
The description in terms of $q$ $\phi_i$ fields turns out to be redundant 
at least as far as the saddle-point solution goes, since
explicit solution for various $q$ reveals that the $\phi_i$
only take on two values in the dominant low-temperature solutions
and all are equal at high temperature\footnote{This is rather
similar to the one-dimensional Potts model where due to the permutation
symmetry (for $q \ge 2$) only two of the $q$ eigenvalues in the transfer matrix 
contribute in the thermodynamic limit.}.
It is thus possible to write down an effective
saddle-point action
in terms
of just two fields $\phi=\phi_{1 \ldots q-1}$ and $\tilde \phi = \phi_q$
\begin{eqnarray}
S_L = {1 \over 2} ( q - 1) \left[ 1 - c ( q - 2) \right] \phi^2  - { 1 \over 3} (
q -1) \phi^3
+ {1 \over 2} \tilde \phi^2  - {1 \over 3} \tilde \phi^3 - c ( q -1 ) \phi \tilde \phi.
\label{app1}
\end{eqnarray}
If we solve for $\phi, \tilde \phi$ we find
the low-temperature solutions 
\begin{eqnarray}
\phi &=& \frac{1}{2} - \frac{1}{2} (q-3) c-\frac{1}{2} \sqrt{(1-2cq+2c+c^2q^2-6c^2q+5c^2)}, \nonumber \\
\tilde \phi &=& \frac{1}{2} + \frac{1}{2} (q-1) c + \frac{1}{2} 
\sqrt{(1-2cq+2c+c^2q^2-6c^2q+5c^2)}
\label{phisols}
\end{eqnarray}
as well as a high-temperature solution with $\phi = \tilde \phi = \phi_0$,
\begin{equation}
\phi_0 = 1 - (q-1) c,
\end{equation}
and another branch of the solution reversing the signs in front of both square roots.
These leads to a high-temperature saddle-point action
\begin{eqnarray}
S_H = \frac{q}{6} ( 1 - (q - 1)c )^3, 
\label{pottsact}
\end{eqnarray}
and two unwieldy expressions
for the
saddle-point action $S_L$ in the low-temperature regime,
one coming from each branch of the $\phi$ solutions,
\begin{eqnarray}
S_L &=& -11/6c^3+1/12q-1/4cq^2-1/12c^3q^4-3/2c^2q^2-13/4c^3q^2
\nonumber \\
&+& 1/4 c^2q^3+9/4c^2q+11/12c^3q^3-c^2+17/4c^3q+3/4cq-1/2c  \nonumber \\
&\pm& 1/12(q-2)(1-2cq+2c+c^2q^2-6c^2q+5c^2)^{3/2}.     
\end{eqnarray}
The expressions for $S_L$ become
a lot more palatable when a particular $q$ value is inserted.
Just as for the Ising model the locus of Fisher zeroes
can now be obtained by demanding $| S_H | = | S_L|$
for complex $c$.

A schematic graph of the locus of zeroes for general $q>2$ is shown
in Fig.~2 where it can be seen that the physical locus of zeroes, shown
in bold, crosses the real axis in the physical region
of the parameter ($0<c< 1 / ( q -1)$) at the first-order transition
point ${\bf Q} = ( 1 - ( q - 1)^{-1/3}) / ( q - 2)$. This physical locus lies
inside a larger metastable locus, shown dashed,  which has a cusp at the 
spinodal point ${\bf P} = 1 / ( 2 q - 1 )$. The loop in the Ising locus,
which was excised by comparing the actions along the real axis, has split
into the two dashed mirror inner loops lying off the axis in the  
Potts models.
Referring back to Fig.~1 we can see that in
the Ising case the first-order
locus vanishes but the cusp remains and takes over the
role of  the physical locus.
At ${\bf Q}$ we would assign $\alpha=1$ for a first order transition
point so the angle of approach to the real axis should be 
$\pi/2$ from
equ.(\ref{slope}), which is what is observed. The other crossing point
${\bf O}$ lies outside the physical region $0<c< 1 / ( q -1)$.

Perhaps a slightly fairer comparison with the Ising model is
to plot the loci in the $a = \exp ( - 2 \beta)$ plane which we have also
done in Fig.~3. The outer ovoid is now the genuine locus of zeroes
and the metastable locus maps onto the tri-foliate structure that lies
largely inside this. In the variable $a$ the physical region is
$0 \le a \le 1$.

\section{Potts model and thin chromatic zeroes}
  
One particularly interesting feature of the Potts saddle-point
solution is that $q$ appears as a parameter, so it is possible
to obtain the Fisher locus for non-integer or even complex $q$-values.
This means that in principle one can also investigate
the complex $q$-properties of the chromatic polynomials on the random graphs.
The chromatic polynomial $P(G,q)$ on a given graph
$G$ of order $n$ encodes the number of ways that $G$
may be $q$-coloured so that no two adjacent vertices have the 
same colour. 
It is related to the partition function of the
anti-ferromagnetic Potts model $Z ( G, q, \beta = -\infty)$ on $G$
in the zero-temperature limit,
\begin{equation}
Z ( G, q, \beta = -\infty) = P( G, q).
\end{equation}
We can follow Shrock and Tsai
\cite{evenmoreshrock}
and consider a limiting function $W ( \{ G \} ,q)$ on some
class of graphs $\{ G \}$, in our case random regular graphs,
\begin{equation}
W ( \{ G \} ,q) = \lim_{n \to \infty} P ( G, q)^{1/n},
\end{equation}
which is related to the free energy (per site) of the corresponding
Potts model
\begin{equation}
f ( G, q, \beta = -\infty) = \log W ( \{ G \} ,q),
\end{equation}
and gives us that the saddle-point action is
effectively the limiting function, $\tilde S( G, q, \beta = -\infty)  
\sim W ( \{ G \} ,q)$. Since $\beta \to  -\infty$ corresponds to
$c \to 1 / ( q - 2)$ we can write down an action in this limit 
by making a suitable substitution for $c$ in equ.~(\ref{app1})
\begin{equation}
S =  - \frac{1}{3} ( q - 1) \phi^3 + \frac{1}{2} \tilde \phi^2 - \frac{1}{3}
\tilde \phi^3 - { q - 1 \over q - 2 } \phi \tilde \phi.
\label{chrom}
\end{equation}
Instead of looking for high- and low-temperature saddle-point solutions of this
action, we are now interested in finding the ``high $q$'' ($S_H$, $\phi_1=\phi_2$) and 
``low $q$'' ($S_L$, $\phi_1 \ne \phi_2$) solutions,
which are given by
\begin{eqnarray}
\phi_0 &=& \phi_1 \; \; = \; \phi_2 \; \; = \; {1 \over 2 - q }, \nonumber \\
\phi_{1,\pm} &=& {-1 \pm \sqrt{5 - 4 q} \over 2 ( 2 - q)}, \\
\phi_{2,\pm} &=& { (2 q - 3) \pm \sqrt{ 5 - 4 q} \over 2 ( q - 2)}, \nonumber 
\end{eqnarray}
respectively.
We can now ape the procedure for finding the locus of Fisher
zeroes to determine the locus of chromatic zeroes by examining
the curves defined by $|S_L| = |S_H|$ in the complex $q$-plane
rather than the complex $c$-plane. Although presented in a 
disguised form this turns out to be the Lima\c con of Pascal
shown in Fig.~4, adding nicely to our collection
of classical plane curves. We can
see that, as expected, there are real zeroes at $q=1$ and $q=2$ indicating that
the three-regular random graph cannot be one or two-coloured. Interestingly 
a zero also appears on the negative real axis, which is a somewhat unusual
occurrence. There is no zero at $q=3$ since
Brook's theorem in graph theory \cite{21}
tells us that if G is a connected graph with maximal co-ordination number
$z>2$,
and if G is not complete,
then the chromatic number $\chi$
of G is $z$.
Here we have a 3-regular graph that is not complete,
so $\chi=3$
and we can avoid frustration completely with 3 colours.

\section{Discussion}

We have seen that both Fisher and chromatic zero  loci may be 
traced out on thin random graphs by adapting the approach of \cite{BK}
which was originally formulated in the context of first order phase transitions.
We argued that even transitions that were continuous at the physical
critical point
would generically display first order properties at all but a finite 
set of points and that the formulae of 
equ.(\ref{master}) could thus still be
employed. These gave a cusp-like locus,
similar to an amputated Folium of Descartes, for the Ising Fisher zeroes and
a rather more complicated structure for the $q>2$ Potts models, where
the Ising-like cusp was still present as a metastable locus. The 
limiting chromatic zero locus on $\phi^3$ graphs, remarkably, emerged as a 
Lima\c con. The Ising model on the Bethe lattice with 3 neighbours per site
returned the same Fisher zeroes as the $\phi^3$ graphs by a rather 
different route, showing again the effective equivalence of the
models.

We have not discussed the case of {\it planar} random graphs here, but 
a similar approach can be applied to the solutions of \cite{boul},
and compared with the series expansions and Monte-Carlo simulations of 
\cite{jan}. Similarly, we have concentrated entirely on Fisher zeroes, but
it is clear following \cite{BK}
that one can also examine the behaviour of the Lee-Yang zeroes.

In summary, noting  that the locus of partition function zeroes
can be thought of as phase boundaries in the complex temperature
plane gives a useful way of determining the 
the locus of Fisher zeroes in the thermodynamic limit 
for Ising and Potts models on random graphs, as well
as chromatic zeroes.

\section{Acknowledgements}

B.D. was partially supported by Enterprise Ireland Basic Research Grant - SC/1998/739 and B.D. and D.J. by an Enterprise Ireland/British 
Council Research Visits Scheme - BC/2000/004. 
B.D. would like to thank Heriot-Watt mathematics department for its 
hospitality and D.J. would like to thank
the Department of Mathematical Physics,
National University of Ireland,
Maynooth for the same. W.J. and D.J. were partially supported by
ARC grant
313-ARC-XII-98/41
and  the EC IHP network
``Discrete Random Geometries: From Solid State Physics to Quantum Gravity''
{\it HPRN-CT-1999-000161}.
\bigskip
%

%
%
\clearpage \newpage
\begin{figure}[htb]
\vskip 20.0truecm
\includegraphics{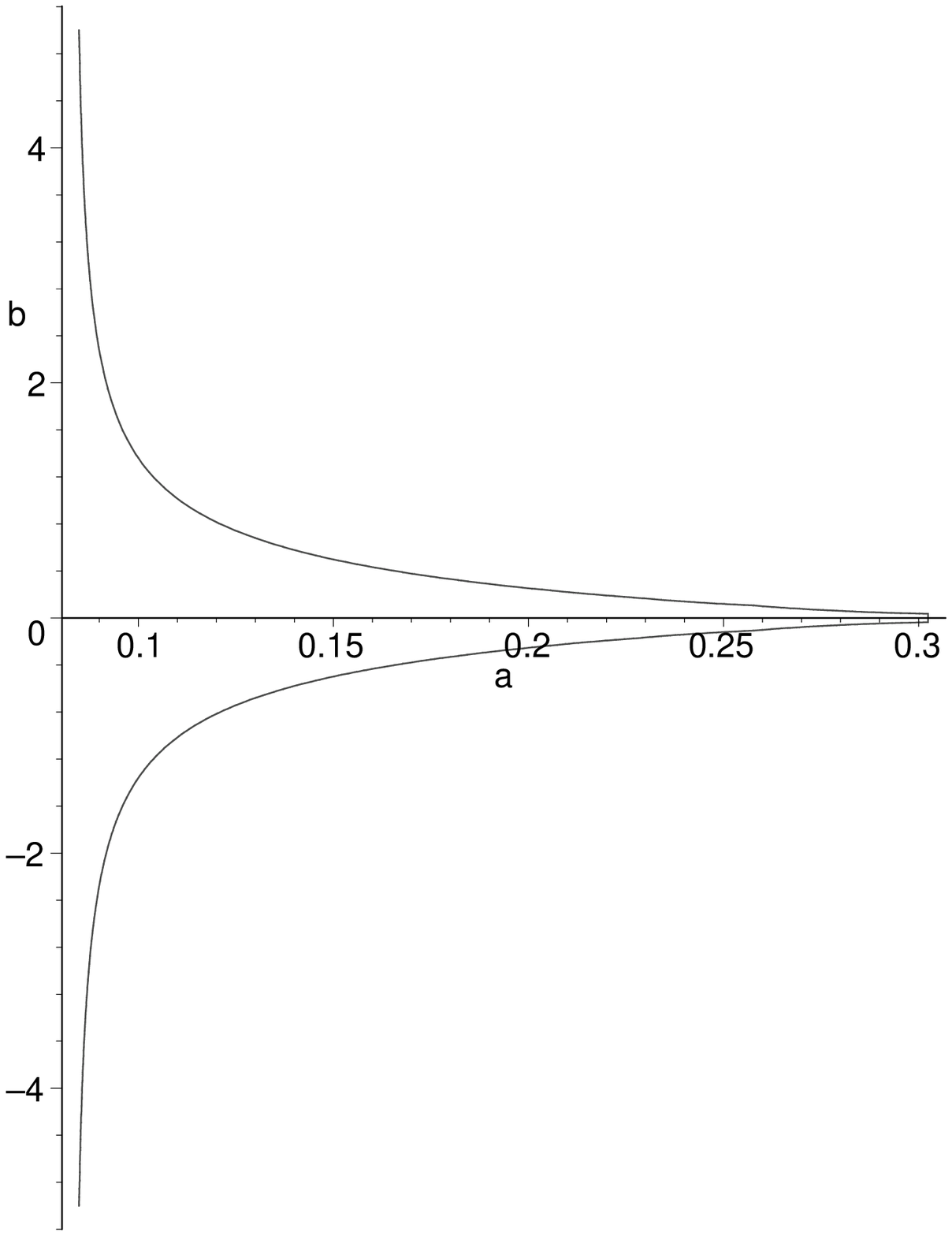}
\caption[]{\label{fig1} The locus of Fisher zeroes in the complex
$c=\exp ( -2 \beta)$ plane for the Ising model on $\phi^3$
random graphs. The real and imaginary parts of $c$ are denoted by $a$ and
$b$ respectively.
}
\end{figure}
\clearpage \newpage
\begin{figure}[htb]
\vskip 20.0truecm
\includegraphics{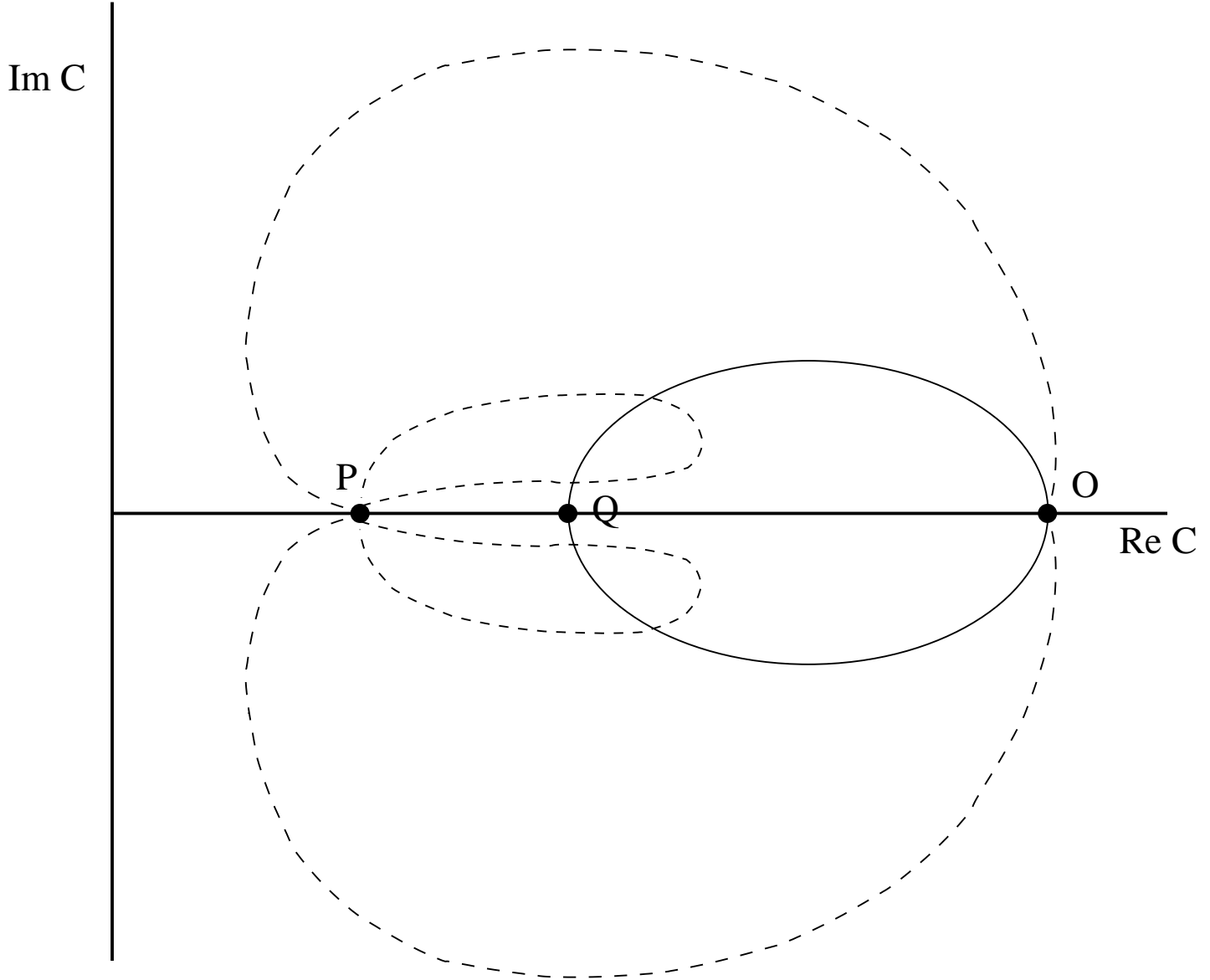}
\caption[]{\label{fig3}
A schematic representation
of the locus of Fisher zeroes for a $q>2$  state Potts model,
in the complex $c  = 1/( \exp (  2 \beta ) + q - 2)$ plane.  
The dashed locus represents the metastable branch and would
{\it not} give rise to partition function zeroes, whereas the
inner loop, drawn in bold represents the true partition function zeroes.
The labelled points are 
the spinodal point ${\bf P} = 1/ (2 q - 1)$, the ``true'' first-order 
point ${\bf Q} = (1 - ( q - 1)^{-1/3}) / ( q -2)$
and the point 
${\bf O}$ at which the loci meet (which is outside the physical region
$0 < c < 1 / ( q - 1)$).
}
\end{figure}  
\clearpage \newpage
\begin{figure}[htb]
\vskip 20.0truecm
\includegraphics{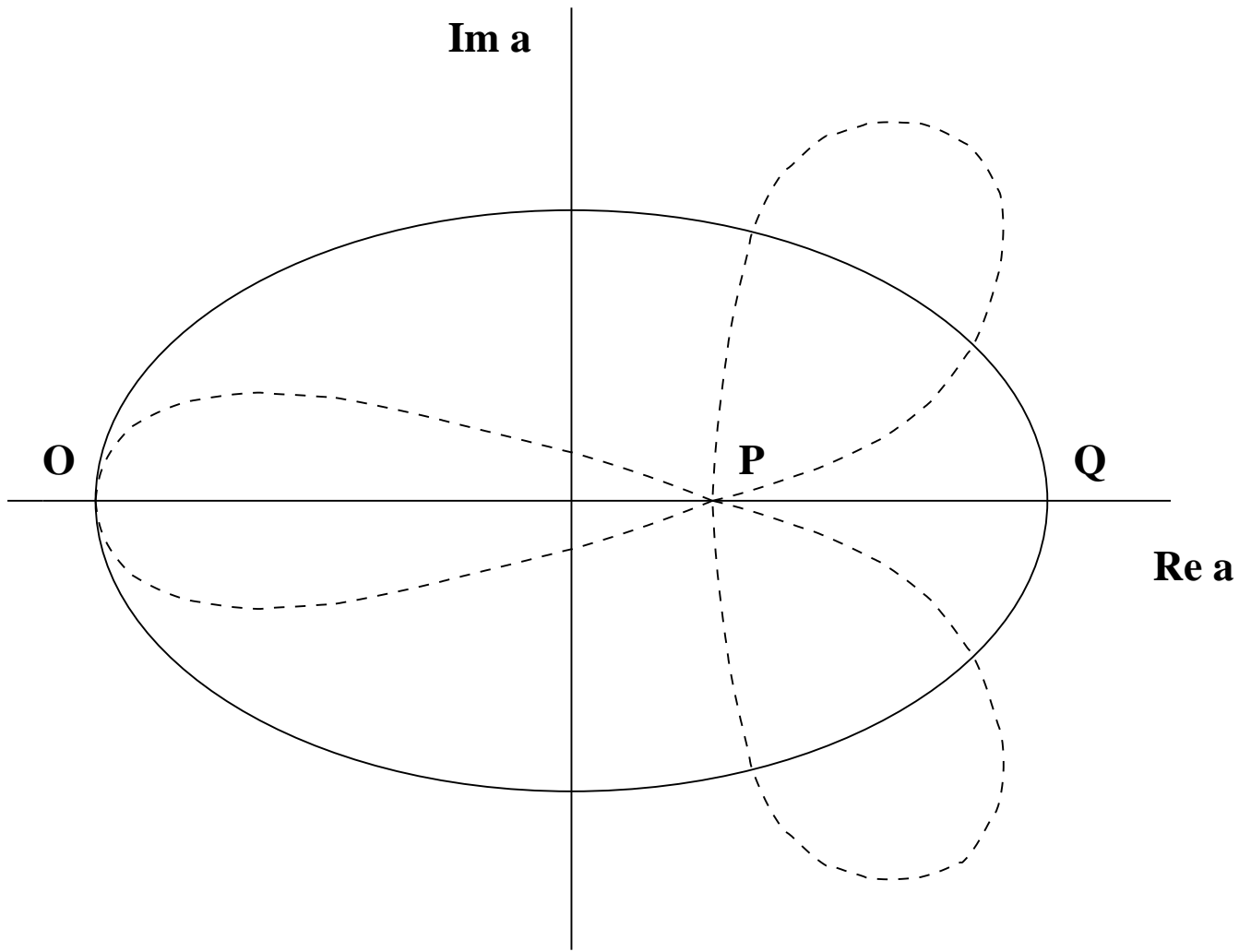}
\caption[]{\label{fig4}
The locus of Fisher zeroes for a $q>2$  state Potts model,
now in the complex $a  = \exp ( - 2 \beta )$ plane. The labelling
is as in Fig.~2.
}
\end{figure}
\clearpage \newpage
\begin{figure}[htb]
\vskip 20.0truecm
\includegraphics{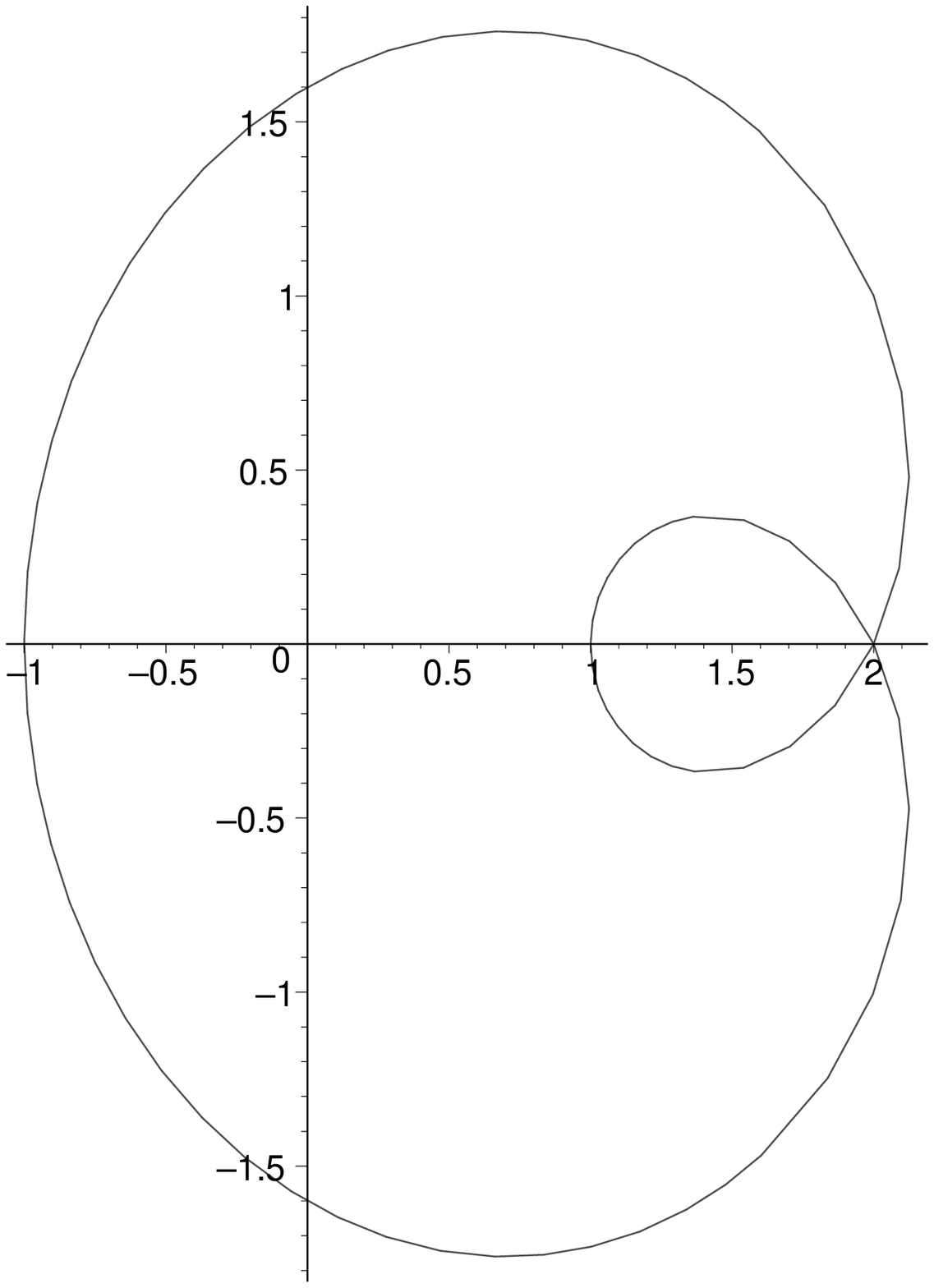}
\caption[]{\label{fig5}
The locus of chromatic zeroes for $\phi^3$ graphs
in the complex $q$-plane. The locus is given by the limacon
$((x-1)^2+y^2-1)^2-(x-2)^2-y^2=0$ where $x=\Re q$ and $y= \Im q$.
}
\end{figure}


\begin{thebibliography}{99}
\bibitem{YL} 
C.N. Yang and T.D. Lee, Phys. Rev. {\bf 87} (1952) 404;\\
T.D. Lee and C.N. Yang, Phys. Rev {\bf 87} (1952) 410.
\bibitem{Fish} J. Lebowitz and O. Penrose, Comm. Math. Phys. {\bf 11} (1968) 99;\\
                 G. Baker, Phys. Rev. Lett. {\bf 20} (1968) 990;\\
                 R. Abe, Prog. Theor. Phys. {\bf 37} (1967) 1070; {\it ibid} {\bf 38} (1967) 72;
                 {\it ibid} {\bf 38} (1967) 568;\\
                 S. Ono, Y. Karaki, M. Suzuki, and C. Kawabata, J. Phys. Soc. Japan {\bf 25} (1968) 54;\\
                 D. Gaunt and G. Baker, Phys. Rev. {\bf B1} (1970) 1184;\\
                 P. Kortman and R. Griffiths, Phys. Rev. Lett. {\bf 27} (1971) 1439;\\
                 M. Fisher, Phys. Rev. Lett. {\bf 40} (1978) 1611;\\
                D. Kurtze and M. Fisher, Phys. Rev. {\bf B20} (1979) 2785;\\
               M. Fisher, in: {\em Lectures in Theoretical Physics\/} {\bf VII C} (University of Colorado Press, Boulder, 1965).
\bibitem{IPZ} C. Itzykson, R. Pearson, and J. Zuber, Nucl. Phys. {\bf B220 [FS8]} (1983) 415.
\bibitem{lottsashrock} V. Matveev and R. Shrock, J.Phys. {\bf A28} (1995) 1557;
J. Phys. {\bf A29} (1996) 803;
J. Phys. {\bf A28} (1995) 4859;
J. Phys. {\bf A28} (1995) 5235; 
J. Phys. {\bf A28} (1995) L533-L539; 
Phys. Lett. {\bf A204} (1995) 353;
Phys. Rev. {\bf E53} (1996) 254;
Phys. Lett. {\bf A215} (1996) 271;
Phys. Rev. {\bf E54} (1996) 6174.
\bibitem{creswick} S.-Y. Kim and R.J. Creswick, Phys. Rev. {\bf E58} (1998) 7006;\\
Phys. Rev. Lett. {\bf 81} (1998) 2000;\\
Physica {\bf A281} (2000) 252;\\
Physica {\bf A281} (2000) 262.
\bibitem{prz} P. Repetowicz, U. Grimm, and M. Schreiber,
Mat. Sci. Eng. {\bf A294-296} (2000) 638.
\bibitem{jan} J. Ambjorn, K. Anagnostopoulos, and U. Magnea, Mod. Phys. Lett. {\bf A12} (1997) 1605;\\
Nucl. Phys. (Proc.Suppl.) {\bf 63} (1998) 751.
\bibitem{brazil} L.C. de Albuquerque, N.A. Alves, and D. Dalmazi,
Nucl. Phys. {\bf B580} (2000) 739.
\bibitem{BK} M. Biskup, C. Borgs, J.T. Chayes, L.J. Kleinwaks, and R. Kotecky,
Phys. Rev. Lett. {\bf 84}
(2000) 4794;\\
and preprint
``Partition function zeros at first-order phase transitions''.
\bibitem{various} 
J.L. Monroe, J. Phys. {\bf A29} (1996) 5421; J. Stat. Phys. {\bf 65} (1991) 255;\\
F. Wagner, D. Grensing, and J. Heide, J. Phys. {\bf A33} (2000)
929;\\
A.Z. Akheyan, N.S. Ananikian, and S.K. Dallakian, Phys. Lett. A
{\bf 242} (1998) 111;\\
N.S. Ananikian, S.K. Dallakian, N.Sh. Izmailian, and K.A.
Oganessyan, Phys. Lett. {\bf A214 } (1996) 205; Erratum {\bf A221 }(1996)
434;\\
N.S. Ananikian, S.K. Dallakian, N.Sh. Izmailian, and K.A.
Oganessyan, Fractals {\bf 5} (1997) 175{\bf ; }N.S. Ananikian and S.K.
Dallakian, Physica {\bf D107} (1997) 75;\\
N.S. Ananikian, S.K. Dallakian, N.Sh. Izmailian, K.A.
Oganessyan, and B. Hu, Phys. Lett. {\bf A248} (1998) 381;\\
B. P. Dolan, Phys. Rev {\bf E52} (1995) 4512-4515, Erratum
{\bf E53} (1996) 6590.
\bibitem{Bac} C. Bachas, C. de Calan, and P.M.S. Petropoulos,
J. Phys. {\bf A27} (1994) 6121.
\bibitem{dj} C.F. Baillie, D.A. Johnston, and J.-P. Kownacki,
Nucl. Phys. {\bf B432} (1994) 551;\\
C.F. Baillie, W. Janke, D.A. Johnston, and P. Plechac,
Nucl. Phys. {\bf B450} (1995) 730.
\bibitem{Glu} Z. Glumac and K. Uzelac, J. Phys. {\bf A27} (1994) 7709;\\
              B. Dolan and D. Johnston, ``1D Potts, Yang-Lee Edges and Chaos'',
              cond-mat/0010372.
\bibitem{Bax} R. Baxter, {\em Exactly Soluble Models in Statistical Mechanics\/} (Academic Press, London, 1982). 
\bibitem{boul} V.A. Kazakov, Phys. Lett. {\bf A119} (1986) 140;\\
             D.V. Boulatov and V.A. Kazakov, Phys. Lett. {\bf B186} (1987) 379;\\
             Z. Burda and J. Jurkiewicz, Acta Physica Polonica {\bf B20} (1989)
949. 
\bibitem{potts} D.A. Johnston and P. Plechac, J. Phys. {\bf A30} (1997) 7349.
\bibitem{evenmoreshrock} R. Shrock and S.-H. Tsai,  Phys. Rev. {\bf E55} (1997)
5165;\\
Phys. Rev. {\bf E56} (1997) 1342; Phys. Rev. {\bf E56} (1997) 3935; Physica 
{\bf A259} (1998) 315.
\bibitem{21} R. Wilson, {\em Introduction to Graph Theory\/}, 3rd edition
(Longman, 1985). 
\end{thebibliography}
\end{document}